\documentclass[12pt,letterpaper]{article}
\usepackage{amsmath,amssymb,pgf,pgfarrows,pgfnodes,float,appendix, hyperref}
\usepackage{graphicx}
\usepackage{subfigure}
\usepackage[margin=0.9in]{geometry}

\title{{\rm\footnotesize \qquad \qquad \qquad \qquad \qquad \ \qquad \qquad \qquad \ \ \ \ \ \                  RUNHETC-2019-15}\vskip.5in    Black Hole Time Scales: Thermalization, Infall and Complexity}
\author{Tom Banks\\
Department of Physics and NHETC\\
Rutgers University, Piscataway, NJ 08854}

\date{}
\begin{document}
\maketitle

\begin{abstract} We argue that the infall time to the singularity in the interior of a black hole, is always related to a classical thermalization time. This indicates that singularities are related to the equilibration of infalling objects with the microstates of the black hole, but only in the sense of classical equilibration.  When the singularity is reached, the quantum state of the black hole, initially a tensor product of the state of the infalling object and that of the black hole, is not yet a "generic" state in the enlarged Hilbert space, so its complexity is not maximal.  We relate these observations to the phenomenon of mirages in the membrane paradigm description of the black hole horizon and to the shrinking of the area of causal diamonds inside the black hole.  The observations are universal and we argue that they give a clue to the nature of the underlying quantum theory of black holes in all types of asymptotic space-times.  
\end{abstract}

\section{Introduction}

The nature of the space-like singularity inside the horizon of black holes has been the subject of conjecture since it was discovered.  Recent progress on the theory of quantum gravity has not led to a resolution of this problem.  One observable of the local GR description of black holes, which is difficult to describe in a formalism in which all variables live on the conformal boundary of an infinite space-time, is the infall time to the singularity.  The infamous {\it firewall problem}\cite{firewall} is essentially the claim that the infall time to the singularity is of order the Planck time.
Much of the recent discussion of firewall avoidance in the context of AdS/CFT has focussed on the quantum mechanical notion of computational complexity.  It's argued that while a generic quantum state has a firewall, one does not create a generic quantum state when a black hole is formed by collapse or collision of ordinary matter.
The time required to reach a generic state is extremely long, and for black holes in Minkowski or dS space, or small black holes in AdS space, it's exponentially longer than the black hole evaporation time.  Indeed, for systems with finite lifetimes much shorter than the time it takes to develop maximal complexity from a given initial state, the whole concept of complexity does not make any sense.

This analysis does not explain the infall time to the singularity, as experienced along a time-like trajectory in the interior of the black hole, which means that it does not answer the only kind of {\it experimental} question one can ever ask about a black hole interior.  Note that black hole interiors never have time-like Killing vectors, so by definition, time evolution generated by an asymptotic symmetry of the space-time in which the black hole is immersed, cannot be directly relevant to any experimental measurement in the interior.  One can of course extend global time slices far from the black hole into the interior in an infinite number of possible ways, but any prescription that avoids the singularity is related to proper time along interior time-like trajectories by a singular transformation.
This is true in particular for the interior slices of the Einstein-Rosen bridge described in \cite{susskind}.

The Holographic Space-time (HST) resolution of the firewall problem has a quite different flavor\cite{firewall3}.  In HST models, as in Matrix Theory\cite{bfss}, a system of two objects far from each other is described by a constrained state of the fundamental variables of quantum gravity.  The Hamiltonian is a trace of a polynomial of matrices that are bilinears in the fundamental variables.  The constraints imply the vanishing of off diagonal matrix elements between two subsystems. In \cite{firewall3} the authors argue that the Hamiltonian describing infall is dual to that describing propagation on the stretched horizon.  From the stretched horizon perspective, one sees equilibration of the two systems due to excitation of the constrained, off-diagonal, degrees of freedom.  This is interpreted along the infalling trajectory as a decrease in entropy available to describe local physics of the infalling system, corresponding to the decrease in holographic screen area of causal diamonds as their past tips are taken further from the horizon.  The states that have come into equilibrium with the pre-collision black hole's variables are no longer included in the Hilbert space that is causally connected to the infalling system. 

It turns out that this connection between infall time to the singularity and equilibration time of the extra entropy that is produced by the collision is quite general and holds in all the examples I have checked.  Let's begin by reviewing what happens in Minkowski space.  The Schwarzschild metric is 
\begin{equation} ds^2 = - (1 - \frac{R_S}{r}) dt^2 + \frac{dr^2}{(1 - \frac{R_S}{r})} + r^2 d\Omega^2 . \end{equation}  Inside the horizon $r$ becomes the time-like coordinate and the proper time for any infalling time-like trajectory to hit the singularity is of order $R_S$.  A system of mass $m$ falling on the black hole increases its entropy according to the first law of thermodynamics
\begin{equation} m /T = dE/T = dS = R_S m . \end{equation} Here and in what follows we set the Planck scale equal to one and neglect order one geometric factors.  The question is, where does that huge increase in entropy come from?  The microscopic object falling on the black hole has entropy that is $o(1)$.   It seems rather inevitable that we must view the system of infalling particle plus black hole as existing initially in a constrained state, in which a number of degrees of freedom (q-bits) of order $m R_S$ are frozen.  The temperature $T$ is the average energy per degree of freedom of the system, and therefore sets the scale for the {\it fastest} time scales on which equilibration can take place.   Note that the huge black hole entropy implies that there are splittings of quantum levels of the black hole that are exponentially smaller than $T$.   Classical and quantum recurrence times are controlled by these level splittings, but they are actually meaningless for this system, because the black hole decays long before one can resolve them.  In particle physics language, the width of the excitation is much larger than the splittings involved in recurrences and the growth of complexity.  The eternal Schwarzschild solution of GR is better thought of as a description of a large black hole onto which a spherically symmetric dribble of infalling matter is constantly falling, at a rate that makes up for the mass loss to Hawking radiation, rather than an isolated quantum system with a fixed Hilbert space.

The above discussion used the proper time of an infalling trajectory to define the frame in which energy and temperature are defined.  It can be compared to the view of an accelerated trajectory outside the black hole, {\it e.g.} 
at a fixed Schwarzschild coordinate $r \sim R_S$.  This defines a correspondence between the times along the two trajectories, which can be synchronized when the infalling trajectory crosses that value of $r$.  The time of horizon crossing for the infalling trajectory can be synchronized by equating it to the time that a signal of fixed frequency emitted from that trajectory, falls below a certain infrared cut-off frequency when received at the position of the accelerating trajectory.  This definition has only a weak dependence on the choices of fiducial and cutoff frequencies.

The quasi-normal ringdown time for quasi-normal modes seen from the accelerating trajectory, is of course of order $R_S$, which is the same as the infall time to the singularity.  The fact that the apparent horizon of a trajectory crossing the real horizon begins to shrink is interpreted as the decrease of the number of degrees of freedom of the system along the infalling trajectory, which are not yet in equilibrium with the black hole.

Hayden and Preskill\cite{hp} argued that the number of units of $R_S$ that it takes to come to classical equilibrium is of order ${\rm ln} (R_S)$ and Susskind and Sekino\cite{ss} argued that this implied that the quantum system responsible for black hole thermodynamics had to be a fast scrambler.  An equivalent statement is that it should be invariant under a finite truncation\footnote{because the total entropy is finite} of the group of volume preserving mappings of the horizon and should not behave like a local field theory.

\subsection{1 + 1 Dimensional Models}

In\cite{tbldbh} I described a Holographic Space-time (HST) derivation of the famous exactly soluble $0B$ string theory.  The HST analysis led to a generalized set of consistent quantum models which had meta-stable excitations with the properties of linear dilaton black holes.  The models had many free parameters and I conjectured that at least some choices of those parameters would correspond to models dual to the near horizon limit of linear dilaton black holes arising in Calabi-Yau compactifications of Type II strings.  The low energy effective field theories of those models are the CGHS\cite{cghs}  models of large numbers of massless fermions coupled to $1 + 1$ dimensional linear dilaton gravity\footnote{$1 + 1$ dimensional gravitational models always contain a scalar field, whose non-trivial ground state profile measures the entropy in causal diamonds with given proper time.  Different models have different ground state profiles. }.  Since every pair of Dirac fermions can be mapped onto the $1 + 1$ dimensional string theory, we conjectured that the UV model containing $2N$ relativistic fermions was related to a model of $N$ non-relativistic fermions in the upside down harmonic oscillator potential, with single body Hamiltonian $h = \frac{1}{2} (p^2 - \lambda^2) . $ \cite{moore}.   The scale of energy is defined by the curvature of the upside down oscillator, which we've set equal to 1,  and we identify this with the two dimensional analog of the Planck scale.  The other scale in the problem is the Fermi level, and scattering amplitudes are small when the Fermi level is far below the top of the potential.  The string coupling is roughly the ratio of the Planck scale to the distance between the Fermi level and the zero of potential energy.  The gravitational problem is one sided and the even and odd wave functions in the potential correspond to excitations of two different relativistic fermion species.  

The variables $u_{\pm} = \frac{1}{\sqrt{2}} (p \pm \lambda) $ satisfy the 't Hooft-Dray commutation relations between incoming and outgoing light front coordinates at a black hole horizon.  The Hamiltonian can be written as $\partial_{{\rm ln}\ u_{\pm}} \pm i $ in the asymptotic regions where one or the other variable is large and the naive S-matrix is just the mapping between the basis of fields in these two variables.  In the asymptotic region, the fields are manifestly massless relativistic fermions\cite{AKK}, as one expects in the CGHS models.  

The 0B string theory failed to have black hole excitations for two unrelated reasons.  There are no states of large entropy localized near the region of strong string coupling, which in $\lambda$ space is the region of small $u_{\pm}$ .  Putting in a large number of fermions helps with this problem, but since the model remains integrable there are at least some initial states where one can insert a large amount of energy into the strong coupling region without creating a high entropy meta-stable state, which decays thermally. To ameliorate this problem we introduce an additional interaction term
\begin{equation} \Delta H = \int d\lambda d\lambda^{\prime} J_{IJKL} \Psi_I^{\dagger} ( \lambda ) f (u_+^2 + u_-^2 ) \Psi_J (\lambda)  \Psi_K^{\dagger} ( \lambda^{\prime} ) f ([u^{\prime}_+]^2 + [u^{\prime}_+]^2 ) \Psi_L (\lambda^{\prime}) . \end{equation}
The function $f$ falls off exponentially for large values of its argument.  The couplings $J_{IJKL}$ should couple each fermion to every other one, so that the system will be a fast scrambler\cite{ss}. They should scale with $N$ like the couplings in the SYK\cite{syk} models, so that there is a consistent large $N$ limit. They should also be sufficiently attractive that the system forms large entropy meta-stable bound states when a large number $K$ of different fermions are thrown into the interaction region.  Note that all the scales in the interaction are of order $1$ so that the equilibration time is just the logarithm of the entropy $K \sim N$ . This is consistent with the fact that the infall time to the singularity of linear dilaton black holes is the Planck scale, independently of the size of the hole.  Note also that, following the large $N$ rule that energies, etc. are taken to be $o(1)$ as a function of $N$, the entropy of a black hole will be linear in its energy with a temperature of order Planck scale.  This coincides with the classical entropy formula for linear dilaton black holes. If the couplings transport energy among the different fermion species on the fast scrambling time scale, then the same will be true for black holes formed by a single species of incoming fermion with energy of order $K$.  Note that the vanishing specific heat of a linear dilaton black hole tells us that there are no frozen degrees of freedom, which need to be excited when a localized excitation falls onto a black hole.   Correspondingly, in the quantum models, all the entropy is carried by matter fields.   The dilaton keeps a hydrodynamic record of where the entropy is localized in meta-stable excitations of the model, but there is no entropy associated with the dilaton/graviton fields themselves.

This is to be contrasted to what happens for meta-stable black holes in higher dimensions. The difference is precisely accounted for by the fact that above three dimensions\footnote{Note that in $AdS_3$ there are no meta-stable black holes, though in realistic AdS/CFT models, with two or more compact dimensions of size equal to the AdS radius, there are meta-stable black holes in $5$ or more dimensions, with radius much smaller than the AdS radius, which behave in many ways like Minkowski black holes. }  there are soft graviton modes in a causal diamond of size $R$, which travel only on the boundary of the causal diamond.  These are responsible for the entropy of the diamond, and have an energy measured along causal time slices in the diamond, which is of order $1/R$.  States with a higher degree of localization in the diamond are constrained states of these degrees of freedom. 

One can also study quasi-normal modes of the scalar fields dual to the massless fermions of the model. The authors of \cite{cghs} wrote down the general solution of linear dilaton gravity coupled to scalars dual to an arbitrary number of fermions.   We want to study the linearized solution corresponding to a black hole with an infinitesimal matter perturbation falling on it.  The exact metric is conformal to Minkowski space and the exact scalar field equation is the massless Klein-Gordon equation.  The conformal factor of the metric and the dilaton are given in terms of two massless solutions of the KG equation $v$ and $w$ by
\begin{equation} e^{-2\phi} = v - h_+ h_- , \end{equation}
\begin{equation} e^{-2\phi} = e^{-w} e^{-2\phi} , \end{equation} where 
\begin{equation} \partial_{\pm} h_{\pm} = L_P^{-1} e^{w_{\pm}} . \end{equation} Finally,
\begin{equation} 2 v_{\pm} = ML_P  - \int e^{w_{\pm}} \int e^{-w_{\pm}} \partial_{\pm} f_i \partial_{\pm} f_i . \end{equation} The $\pm$ subscripts denote right and left moving parts of the massless KG solutions. 
To linearized order, the matter perturbations simply propagate into the black hole geometry in a time of order Planck scale, leaving no trace on the horizon.  The quadratic effect of the perturbations is a shift in the horizon.  This is not of course surprising, given that the horizon is a point in this model.   All the perturbation can do is shift the value of the dilaton at the horizon, which is to say, the mass of the black hole.  

Finally, let us digress to note a feature of these models that throws some light on the nature of  string perturbation theory.  One gets into a weakly coupled regime by expanding in powers of the ratio between the Planck scale and the depth of the Fermi surface below the top of the potential.  In this regime, the scattering is universal and integrable.  If we start from a state in which all the individual fermion numbers have fixed values, then to all orders in perturbation theory we just reproduce $N$ commuting copies of the $0B$ S-matrix.  The series are of course not Borel summable.  We could define the model by the fermionic model with any value of the couplings $J_{IJKL}$ and any function $f$ with the same support properties.  All of the UV completions we've suggested are unitary, and consistent with causality.   Many, but not all, of them have meta-stable excitations with the properties of the linear dilaton gravity black holes and deserve to be called quantum models of linear dilaton gravity.  The string perturbation series cannot distinguish between the models with black holes, and those without them.

It's of course possible that if we started from the higher dimensional string models, which have black holes whose near horizon region is described approximately by linear dilaton gravity coupled to massless fermions, we would find string scale corrections to the effective action, which would indicate the need for more elaborate string perturbation series and violate integrability.  However, any such framework would have to have an analog of the tachyon wall, which kept particles of string scale energy away from the top of the potential to all orders in perturbation theory.  It still could not probe the Planck scale interactions responsible for black hole physics in these models.

\section{Stable Black Holes in AdS space}

In studying thermalization and infall for excitations of stable black holes in AdS space, we have to keep careful track of the various scales involved.  We have the AdS radius, the Planck scale, the size of the black hole and the ratio of the size of the infalling disturbance to the AdS radius.   We might also have the string scale in weakly coupled string models, but we will ignore this complication.    

Invoking the AdS/CFT correspondence, we of course want to study a CFT with a gravity dual and black holes whose radius is large compared to $R_{AdS}$. We want to study excitations of the black hole confined to the interior of a cone whose opening angle is small compared to a full sphere.  From the CFT point of view, the whole system lives on a sphere of radius $R_{AdS}$ and opening angles are conformally invariant so our cone has a small opening angle on the boundary sphere.  

There are two time scales associated with scrambling in a general quantum system.  Following \cite{mss} we call these the dissipation time $t_d$ and the scrambling time $t^*$.  For a CFT with a gravity dual, $t_d \sim \beta$ is computed in terms of the decay time of quasi-normal modes for large black holes\cite{hh}, while the results of\cite{mssrefs} indicate that $t^* / t_d \sim {\rm ln S} $\footnote{These estimates are not sensitive enough to determine whether the entropy in the logarithm is that of the entire black hole, or only a patch of size $R_{AdS}$ on the AdS $d - 2$ sphere.}  For an angularly localized perturbation, there is a more refined question one can ask, namely how long it takes the information in the perturbation (i.e. the small number of operators affected by it), to spread over the sphere.   Causality of the CFT tells us that this time, call it $t_{hom}$ is bounded from below by something of order $t_{hom} \sim t_d \beta^{-1} R_{AdS} \sim R_{AdS}$.  That is, scrambling is ballistic in field theory, so one cannot cover the distance around the sphere in time $\beta$.   

All of these times are measured in the field theory on the boundary sphere.  However, the angular opening of the perturbation cone is a conformally invariant quantity, and for a large radius black hole, the metric rapidly approaches the asymptotic form of the AdS metric, in which distances on the sphere and time scale the same way, a few Schwarschild radii from the horizon.  Thus, the time $t_{hom} \sim R_{AdS} $ is the 
time it takes for disturbances to homogenize on the black hole horizon, as seen from a stable orbit around the black hole.  Now recall that the spacetime near the horizon of a large radius AdS black hole has the same constant negative curvature as the background AdS space, and also that proper time it takes to traverse a radially infalling time-like trajectory is never larger than $R_{AdS}$.  Thus, the infall time to the singularity is of order $t_{hom}$.  

What then is the meaning of $t_d$ and $t*$ for an angularly localized perturbation?  Here we have to recall that in all well established examples of AdS/CFT, a large radius AdS space is always accompanied by two or more compact dimensions with approximately the same radius of curvature.  Furthermore the dynamics is not in any way local in the compact dimensions.  In Lagrangian examples like ${\cal N} = 4$ SYM, the angular dynamics on the internal sphere is controlled by a matrix model and is a fast scrambler.  Perturbations localized on the boundary sphere of AdS and on $S^5$, are generically not localized on $S^5$ after a time $t^*$, but remain fairly localized in the transverse AdS directions.   Furthermore, it's clear from examples that a finite fraction of the entropy of the system comes from states associated with changes in the compact dimensions.  Thus, a perturbation to an operator affecting only a few qubits of the system will rapidly spread among all energetically accessible operators that change the compact dimensions, without at first changing its angular location on the horizon.  This means that the space-time picture of the black hole interior in the AdS dimensions, is not greatly affected.  Space-time locality on the sphere in AdS space is not completely destroyed because the tensor network cutoff of the CFT dynamics on the horizon is a lattice field theory with an AdS scale lattice spacing.  The timelike trajectory entering the sphere at fixed angle survives as the description of a node on the lattice, with many degrees of freedom, until ballistic propagation of information has spread the information originally encoded in that node into distant nodes on the sphere.

If, in the geometrical description of the interior, several perturbations are sent in at closely spaced angles, in such a way that the corresponding timelike trajectories are in causal contact before hitting the singularity, then the corresponding CFT description is of two close by nodes interacting, before the information encoded in them propagates around the horizon.  Since we're considering models in which the AdS radius is large, there is time for lots of complex interaction to take place.   

In \cite{firewall3} the authors argued that the well known mirage\footnote{An example of such a mirage: if a particle and anti-particle are sent into a black hole on trajectories such that they meet and annihilate before hitting the singularity, then the field configuration on the horizon has a dipole, which shrinks and disappears before the system reaches equilibrium.} phenomena on the black hole horizon in Minkowski space were equivalent to the dynamics described on trajectories falling through the horizon.  The arguments of this note extend that correspondence to the case of large AdS black holes, though it would probably be worth exploring AdS mirages and their correspondence with the dynamics of the interior in more detail.   For the $1 + 1$ dimensional linear dilaton models, the classical estimate of the infall time is the Planck scale, independent of the black hole size, so there is no such thing as a classical black hole interior.  Indeed, all of the dynamics associated with these very simple black holes in our explicit quantum models takes place at the Planck scale, and is invisible even to string perturbation theory.

\section{Conclusions}

We've argued that the time scale for a geodesic to hit the singularity for a large class of black holes, is of order the time for information to homogenize on the spherical horizon, with the latter time measured along time-like trajectories a few Schwarzschild radii from the horizon, {\it e.g.} stable orbits for massive particles.  In more picturesque language, we've argued for a duality between the "mirage on the horizon" of an event in the interior of the black hole, and the event itself.  The black hole singularity, which in dimensions above three, corresponds to a shrinking of the area of causal diamonds as their past tip approaches the singularity, is associated with the equilibration of degrees of freedom corresponding to subsystems that are weakly coupled to each other (and therefore localized) , with the majority of the degrees of freedom on the black hole horizon.  

All of the time scales discussed in this note are exponentially shorter than the time scales required for the initial quantum state to become generic in the space of all states in "the Hilbert space of the black hole".  Indeed, for black holes in de Sitter or Minkowski space, or small meta-stable black holes in AdS space, the phrase in quotation marks only makes sense on time scales much smaller than the time required for the quantum state to be fully randomized. Thus, issues of complexity have nothing to do with the singularities inside black holes.   It has been conjectured\cite{susskind} that firewall singularities, which destroy the smooth interior geometry of black holes within a Planck time of horizon crossing, are connected with the black hole having reached a state of maximal complexity. This conjecture only makes sense for stable AdS black holes, which actually last for quantum recurrence times, but it is not clear that it is correct even in that case.

The picture of the black hole interior for black holes with negative specific heat, which was developed in \cite{firewall3}, is quite different.  For these black holes, the non-singular interior of a truly isolated black hole, once it has come to thermal equilibrium, does not exist.  Rather, it is recreated each time a light object falls on the black hole, and is dual to the process of thermalization of the frozen degrees of freedom, which mediate interaction between the black hole horizon and the localized object.  The non-singular region of the eternal black hole solution of GR is a feature of a black hole in unstable thermodynamic equilibrium with an infalling gas of particles, at a rate that compensates for Hawking evaporation.  For linear dilaton black holes, there is no non-singular region of super-Planckian extent, and this corresponds to the fact that equilibration takes place on Planck time scales in the dual fermion model.  

One can ask whether these ideas extend to large AdS black holes.  The discussion above says that they do, for perturbations of the black hole localized in small angular regions on the sphere in the approximately flat space one obtains for large AdS radius.  This space has at least two more large dimensions than the AdS space itself, in all known models. In the view of the dynamics from a trajectory a few Schwarzschild radii away from the horizon, thermalization of the degrees of freedom corresponding to the geometry of the compact space transverse to $AdS$ happens rapidly, in a time of order the inverse temperature, much shorter than the AdS radius.  However, infalling objects separated by large angles still behave like localized objects in $AdS$, until a time of order the AdS radius.   The dynamical reason for the persistence of localized behavior on the sphere in AdS is quite different from that for evaporating black holes.  Locality persists for a time $t_{hom} > t^*$ becaus the CFT dynamics is explicitly local on the sphere.  It is not invariant under volume preserving mappings, and therefore gives ballistic, rather than fast, scrambling.  For small black holes in large radius AdS spaces, or large black holes in Minkowski/dS space, the persistence of localization on the horizon and in the interior are associated with the negative specific heat of the black hole and the fact that an initial state in which an object falls through the horizon is a constrained state in the Hilbert space of the eventual black hole that is formed by the infall, with an entropy deficit given, for black hole mass much larger than that of the infalling object, by $\sim m (M)^{\frac{1}{d - 3}} $ in Planck units.  The frozen degrees of freedom are precisely the ones responsible for interaction between those of the infalling object and those of the pre-collision black hole.  The time required to equilibrate these variables is the time over which locality is a sensible concept in the interior.

The situation is somewhat different if the infalling objects have fixed and low angular momentum on the compact sphere\footnote{Presumably the same is true for low lying Kaluza Klein modes in any compact space ${\cal K}$ that forms part of an $AdS_d \times {\cal K}$ model with both radii of curvature large.}.  For these there is no need for thermalization of the internal degrees of freedom in order to homogenize the perturbation on the compact sphere.  Nonetheless, the equation $dE = T dS$ is valid for these perturbations, so that the thermalization time is again related to the extra entropy that's created by the infall.  The horizon state of a large Schwarzschild black hole has vanishing mean angular momentum, but is wandering through a Hilbert space built from states with large values of angular momentum.  Thus the infalling angular momentum can get transferred to horizon DOF, entangling the state of the infalling object with the horizon.  
This thermalization process is localized in a small angular region in the boundary field theory, which is equivalent to the same angular region on the black hole horizon, seen from a nearby timelike trajectory. Thermalization occurs ballistically on the sphere, in $AdS_d$ because the underlying dynamics is that of a local quantum field theory with a tensor network/lattice cutoff.  
This implies that angular locality will be preserved for a time of order $R_{AdS}$.   Note that this can be the proper time along any time-like trajectory at fixed global coordinate $r\gg R_S$, since global time and the radii of spheres are both multiplied by $r^2$.  The translation of this into the frame of an infalling object is that localization in $AdS_d$ persists for a time of order $R_{AdS}$, after which all local information has "fallen into the singularity" $\equiv$ "thermalized with the large system on the horizon".  The latter equivalence appears non-local to a bulk field theorist and is an indication that the underlying holographic model is NOT well described by local field theory.  The local description of the interior of all black holes is a "mirage", whose validity has a finite lifetime.  

Now imagine that we've let a large AdS black hole evolve to a state of maximal complexity, and let us ask what has to be changed in the above description.  I would claim that the answer is "nothing".  The basis of that claim is that from the point of view of general quantum systems, the current note has been about what might be called classical thermalization and its association with singularities.  It's been recognized for many years that the prescription of computing expectation values in an ensemble of all states, subject to a few constraints on macroscopic averages, is valid only for a special class of operators, whose matrix elements in the energy basis obey simple inequalities\cite{deutschberrysrednicki}, even when the Hamiltonian is chosen from a random matrix ensemble.  One argues that this class of operators includes all reasonable measurements one could make on a large chaotic system.  In field theory, where one generally works in terms of n-point correlations of a finite number of fundamental fields, the statement would be that one expects to use this prescription only to calculate a finite number of n-point functions in the infinite system.  This is the reason that we see thermal results long before the system has reached a state of maximal complexity, or experienced a single classical recurrence.  I find it hard to believe that the measurement corresponding to the existence or non-existence of a firewall, which involves whether the macroscopic properties of an infalling semi-classical system are disturbed or not, is sensitive to the details of the actual state of the system, which differentiate between a truly random state, and one in which "simple" operators have their thermal expectation values.  In other words, the projection operator on firewall states should itself be a "simple" operator.

The second reason to believe that complexity has little to do with drama at the horizon, in the sense that phrase is used in the AMPS paper, is the fragility of complexity.  This is by now well understood, but I believe I first pointed it out to L. Susskind in a lunch conversation with L. Susskind and Adam Brown.  The point is that the event of dropping something onto a black hole always involves increasing the entropy of the system by an amount $dS = dE/T$ .  The minimal entropy change involves an energy change $dE = T {\rm ln} 2$, though it's probably incorrect to use thermodynamics for such a small change in energy\footnote{Note by the way that for an energy even modestly larger than $T$, which could certainly be carried by a single massless particle, the increase in entropy of the black hole is much larger than the particle's internal entropy.  This indicates that even for large black holes in AdS space, the infall catalyzes the unfreezing of degrees of freedom which must be non-locally distributed between the infalling object and the black hole.} The point however is that even this minimal change decreases the complexity, relative to its maximum in the enlarged Hilbert space by a factor that is exponential in the number of q-bits in the black hole.  The time to reach maximal complexity after infall is thus enormous.  So far this is consistent with explanations of the lack of drama at the horizon that attribute it to the failure of the black hole state to be maximally complex.  No matter what the initial state of the black hole, the hole plus infalling system is in a state of its full Hilbert space where complexity is far from maximal.   The real problem with this explanation of lack of drama at the horizon is that it implicitly suggests that there is no drama on time scales shorter than the time for buildup of complexity, but we know that the proper times of infall to the singularity are all exponentially shorter than that.   By contrast, the picture of infall proposed in this note is based on the notion that black hole singularities are dual to complete thermalization of temporarily isolated (and therefore bulk localized) degrees of freedom inside the horizon, with the much more numerous horizon states.  By and large, the time scales for infall to the singularity are captured by the thermodynamic equation $dE = T dS$.  The timescale of infall to the singularity is the time to equilibrate the extra entropy brought into the system by the infalling object, which must perforce have been "frozen" when the infalling object was far from the black hole. The case of large AdS black holes is anomalous, because $AdS_d$ dynamics outside the horizon is actually local in angle.  Thus, the mirage of local dynamics can be upheld for timescales long compared to the equilibration time of the infalling system with its local environment on the $d - 2$ sphere.  

The work of \cite{susskind} has revealed a fascinating connection between the growth of complexity and the growth of the Einstein-Rosen bridge (ERB), a geometrical object in certain eternal black hole solutions.  Originally that work was inspired by the firewall problem, and in particular by the demonstration by Stanford and Shenker\cite{ss2} that some of the pure states in the thermal CFT ensemble corresponding to a stable AdS black hole, had high energy shock waves perturbing the smooth geometry dual to the thermofield double state.  The ERB always grew so that a timelike trajectory crossing the horizon did not encounter those shock waves before it hit the singularity.  The growth of complexity is bounded for a finite dimensional Hilbert space so it was conjectured that a maximally complex state would have a firewall at the horizon.  Note however that the time coordinate in which growth of the bridge comes to an end, is related to the proper time of interior time-like trajectories by a transformation that is singular long before the trajectory hits the space-like singularity of the black hole.  Outside the horizon, the ERB time becomes that of an accelerated timelike trajectory, along which the black hole horizon is seen to equilibrate to a spherical black hole of larger mass on a time scale exponentially shorter than those identified with the growth of complexity. This is dual/complementary to the fact that the time slices on which complexity is described in the interior, never reach the black hole singularity.

The eternal black hole in Minkowski/dS space is clearly not a pure state in the Hilbert space.  Since it's an unstable equilibrium it corresponds to an ensemble of states, in each of which a distribution of infalling matter is dropped onto the hole, at a rate that exactly compensates the mass loss due to Hawking radiation.  Thus, the quantum state of the localized Minkowski or dS black hole NEVER approaches maximal complexity.  It is constantly re-entangled with the quantum states of the infalling systems which sustain its precarious equilibrium.  It seems incorrect to claim that the growth of complexity is ever going to stop for this system.  It's time evolution depends on the microscopic details of the quantum state of infalling matter, which are not encoded in the coarse grained description given by the black hole geometry.  I believe that the same is true for the eternal AdS black hole on time scales long compared to the AdS radius.  Again, if there's any sense to talk about a quantum system associated with the localized horizon, then that system must be in a quantum state that's entangled with the rest of the boundary degrees of freedom, with which it exchanges Hawking radiation many times in the period that it takes complexity to grow in a Hilbert space of fixed dimension equal to the exponential of the black hole entropy.  The "local Hilbert space of states near the horizon" is time dependent and its dimension is constantly shrinking and growing.

The linear dilaton black holes are also, in a sense, a counterexample to the conjecture that firewalls are associated with maximal complexity.  These black holes have a classical firewall: the classical geometry in the interior of the hole becomes singular in a proper time of order Planck scale along any interior timelike trajectory.   The quantum models of these holes proposed in\cite{tbldbh} reproduces this result, but the time to hit the singularity is associated with $t^*$ rather than a complexity time scale of the quantum model, which is exponentially large in the number of fermion fields.  

The connection between the geometry of the ERB and quantum complexity is fascinating and deserves to be understood better, but I do not believe it's related to singularities or the resolution of the firewall paradox.

\begin{center}
{\bf Acknowledgments }\\
The work of T.Banks is {\bf\it NOT} supported by the Department of Energy, the NSF, the Simons Foundation, the Templeton Foundation or FQXi. \end{center}


\begin{thebibliography}{99}
\bibitem{firewall}   N.~Itzhaki,``Is the black hole complementarity principle really necessary?,''
  hep-th/9607028;
   S.~L.~Braunstein, S.~Pirandola and K.~?yczkowski,
  ``Better Late than Never: Information Retrieval from Black Holes,''
  Phys.\ Rev.\ Lett.\  {\bf 110}, no. 10, 101301 (2013)
  doi:10.1103/PhysRevLett.110.101301
  [arXiv:0907.1190 [quant-ph]].
  A.~Almheiri, D.~Marolf, J.~Polchinski and J.~Sully,
  ``Black Holes: Complementarity or Firewalls?,''
  JHEP {\bf 1302}, 062 (2013)
  doi:10.1007/JHEP02(2013)062
  [arXiv:1207.3123 [hep-th]].
\bibitem{firewall3} T.~Banks, W.~Fischler, S.~Kundu and J.~F.~Pedraza,
  ``Holographic Space-time and Black Holes: Mirages As Alternate Reality,''
  arXiv:1401.3341 [hep-th].
\bibitem{bfss}T.~Banks, W.~Fischler, S.~H.~Shenker and L.~Susskind, ``M theory as a matrix model: A Conjecture,'' Phys.\ Rev.\ D {\bf 55}, 5112 (1997) doi:10.1103/PhysRevD.55.5112
  [hep-th/9610043].
  
\bibitem{susskind} L.~Susskind,
  ``Three Lectures on Complexity and Black Holes,''
  arXiv:1810.11563 [hep-th].
  \bibitem{hp} P.~Hayden and J.~Preskill,
  ``Black holes as mirrors: Quantum information in random subsystems,''
  JHEP {\bf 0709}, 120 (2007)
  doi:10.1088/1126-6708/2007/09/120
  [arXiv:0708.4025 [hep-th]].
  \bibitem{ss} Y.~Sekino and L.~Susskind,
  ``Fast Scramblers,''
  JHEP {\bf 0810}, 065 (2008)
  doi:10.1088/1126-6708/2008/10/065
  [arXiv:0808.2096 [hep-th]].
  \bibitem{mss}  J.~Maldacena, S.~H.~Shenker and D.~Stanford,
  ``A bound on chaos,''
  JHEP {\bf 1608}, 106 (2016)
  doi:10.1007/JHEP08(2016)106
  [arXiv:1503.01409 [hep-th]].
    \bibitem{tbldbh} T.~Banks,
  ``Holographic Space-time Models in $1 + 1$ Dimensions,''
  arXiv:1506.05777 [hep-th].
  \bibitem{cghs}  C.~G.~Callan, Jr., S.~B.~Giddings, J.~A.~Harvey and A.~Strominger,
  ``Evanescent black holes,''
  Phys.\ Rev.\ D {\bf 45}, no. 4, R1005 (1992)
  doi:10.1103/PhysRevD.45.R1005
  [hep-th/9111056].
  \bibitem{AKK}   S.~Y.~Alexandrov, V.~A.~Kazakov and I.~K.~Kostov,
  ``2-D string theory as normal matrix model,''
  Nucl.\ Phys.\ B {\bf 667}, 90 (2003)
  doi:10.1016/S0550-3213(03)00546-7
  [hep-th/0302106].
\bibitem{moore}  G.~W.~Moore,
  ``Double scaled field theory at c = 1,''
  Nucl.\ Phys.\ B {\bf 368}, 557 (1992).
  doi:10.1016/0550-3213(92)90214-V
  
  \bibitem{syk}  S.~Sachdev and J.~Ye,
  ``Gapless spin fluid ground state in a random, quantum Heisenberg magnet,''
  Phys.\ Rev.\ Lett.\  {\bf 70}, 3339 (1993)
  doi:10.1103/PhysRevLett.70.3339
  [cond-mat/9212030];
  A. Kitaev, KITP lectures March 2015, Santa Barbara.
  \bibitem{16,17,18,19,5mss} S. H. Shenker and D. Stanford, ÒBlack holes and the butterfly effect,Ó JHEP 1403, 067 (2014) [arXiv:1306.0622 [hep-th]]; S. H. Shenker and D. Stanford, ÒMultiple Shocks,Ó JHEP 1412, 046 (2014) [arXiv:1312.3296 [hep-th]]; A. Kitaev, ÒHidden Correlations in the Hawking Radiation and Thermal Noise,Ó talk given at Fundamental Physics Prize Symposium, Nov. 10, 2014.
Stanford SITP seminars, Nov. 11 and Dec. 18, 2014; S. H. Shenker and D. Stanford, ÒStringy effects in scrambling,Ó arXiv:1412.6087 [hep- th]; D. A. Roberts, D. Stanford and L. Susskind, ÒLocalized shocks,Ó arXiv:1409.8180 [hep-th].
   \bibitem{ss2} S. H. Shenker and D. Stanford, ÒBlack holes and the butterfly effect,Ó JHEP 1403, 067 (2014) [arXiv:1306.0622 [hep-th]]; S. H. Shenker and D. Stanford, ÒMultiple Shocks,Ó JHEP 1412, 046 (2014) [arXiv:1312.3296 [hep-th]].
    \bibitem{hh}  G.~T.~Horowitz and V.~E.~Hubeny,
  ``Quasinormal modes of AdS black holes and the approach to thermal equilibrium,''
  Phys.\ Rev.\ D {\bf 62}, 024027 (2000)
  doi:10.1103/PhysRevD.62.024027
  [hep-th/9909056].
  \bibitem{deutschberrysrednicki}  M. Berry,  J.Phys.,A10,2083, (1977);
J. M. Deutsch,  Phys.Rev.,A43,2046,(1991);
M.~Srednicki,
  ``Thermal fluctuations in quantized chaotic systems,''
  J.\ Phys.\ A {\bf 29}, L75 (1996)
  doi:10.1088/0305-4470/29/4/003
  [chao-dyn/9511001].
  \bibitem{mssrefs} S. H. Shenker and D. Stanford, ÒBlack holes and the butterfly effect,Ó JHEP 1403, 067 (2014) [arXiv:1306.0622 [hep-th]]; S. H. Shenker and D. Stanford, ÒMultiple Shocks,Ó JHEP 1412, 046 (2014) [arXiv:1312.3296 [hep-th]]; 
A. Kitaev, ÒHidden Correlations in the Hawking Radiation and Thermal Noise,Ó talk given at Fundamental Physics Prize Symposium, Nov. 10, 2014.
Stanford SITP seminars, Nov. 11 and Dec. 18, 2014; S. H. Shenker and D. Stanford, ÒStringy effects in scrambling,Ó arXiv:1412.6087 [hep- th]; D. A. Roberts, D. Stanford and L. Susskind, ÒLocalized shocks,Ó arXiv:1409.8180 [hep-th].



\end{thebibliography}
\end{document}